\documentclass[useAMS,usegraphicx,usenatbib]{mn2e}

\title[HST-Scale 3D simulations of MHD disc winds]{HST-Scale 3D simulations of MHD disc winds :  A rotating two-component jet
structure}
\author[Staff et al.]
{J. E. Staff, $^{1,2}$\thanks{E-mail: jan.staff@mq.edu.au}, N. Koning$^3$, R.
 Ouyed$^3$, A. Thompson$^3$, R. E. Pudritz$^{4,5}$\\
$^1$Department of Physics and Astronomy, Macquarie University  NSW  2109,
Australia\\
$^2$Department of Physics and Astronomy, Louisiana State University,
202 Nicholson Hall, Tower Dr., Baton Rouge, LA 70803-4001, 
USA\\
$^3$Department of Physics and Astronomy, University of Calgary,
2500 University Drive NW, Calgary, Alberta, T2N 1N4 Canada\\
$^4$Department of Physics and Astronomy, McMaster University, Hamilton ON
L8S 4M1, Canada\\
$^5$Origins Institute, ABB 241, McMaster University, Hamilton ON L8S 4M1,
Canada}
\begin{document}

\maketitle

\begin{abstract}

We present the results  of  large scale, three-dimensional
magneto-hydrodynamics simulations of disc-winds for different initial
magnetic field configurations.  The jets are followed from the source to 90
AU scale, which covers several pixels of HST images of nearby protostellar
jets.  Our simulations show that jets are heated along their length by many
shocks.  We compute the emission lines that are produced, and find excellent
agreement with observations.  The jet width is found to be between 20 and 30
AU while the maximum velocities perpendicular to the jet is found to be up
to above $100 {\rm km/s}$.  The initially less open magnetic field
configuration simulations results in a wider, two-component jet; a
cylindrically shaped outer jet surrounding a narrow and much faster, inner
jet.  These simulations preserve the underlying Keplerian rotation profile
of the inner jet to large distances from the source.  However, for the
initially most open magnetic field configuration the kink mode creates a
narrow
corkscrew-like jet without a clear Keplerian rotation profile and even
regions where we observe rotation opposite to the disc (counter-rotating).
The RW Aur jet is narrow, indicating that the disc field in that case
is very open meaning the jet can contain a counter-rotating component that we
suggests explains why observations of rotation in this jet has given
confusing results. 
Thus magnetized disc winds from underlying Keplerian discs can develop
rotation profiles far down the jet that are not Keplerian.

\end{abstract}


\section{Introduction}

Astrophysical jets are observed in a range of
astrophysical objects including protostars, X-ray binaries, active galactic
nuclei (AGN), and more.  Common to many of these sites is a central object surrounded by an accretion disc.  The magnetocentrifugal mechanism described
in \citet{bp82} explains how jets can be launched, accelerated, and
collimated by a large scale magnetic field threading the disc.
While this mechanism was originally developed in the context of AGN jets,
its applicability to ubiquitous protostellar jets has been substantially
developed for over 3 decades through detailed theoretical models
\citep[e.g.][]{pelletier92,ferreira97,krasnopolsky02}, advanced simulations in 2 dimensions
\citep[e.g.][]{uchida85,shibata85,ustyugova95,ouyed97,kato02,ramsey11} and 3 dimensions
\citep[e.g.][]{ouyed03,anderson06,moll09,staff10}, and
tested against the ever increasing high resolution observations
\citep[e.g.][]{bacciotti00,bacciotti02,woitas05,ray07,coffey08,coffey12}.
The magneto-hydrodynamic disc wind is now arguably the most
extensively tested theory available for protostellar jets and outflows
\citep[see reviews by e.g.][]{pudritz07,mckee07,frank14}.

Jets from young stellar objects have been seen from both low-mass
protostars, for instance the RW Aur jet studied in \citet{coffey12} and
\citet{woitas05}, and also from massive protostars \citep[for instance G35.2
observed by][]{zhang13}.  \citet{cesaroni13} noticed similarities between
the observed jet from the massive protostar IRAS 20126+4104 and the
simulated jets shown in \citet{staff10}.  While it may well be that the
magnetocentrifugal mechanism can explain jets from both low-mass and massive
protostars, as well as other jets like those in AGNs \citep[see for
instance][]{staff13}, in this paper we focus our attention on jets from 
low-mass (i.e. solar mass) protostars.

The magnetocentrifugal mechanism 
is not the only
proposed theoretical model for young stellar object (YSO) jet formation.  An
alternative model is the X-wind model \citep[e.g.][]{shu00} in which a
Keplerian rotation profile of the jet is not expected.  The two main goals
of the simulations presented here is to look for rotation and magnetic
structures in jets that lead to features reminiscent of observed jets, and
that can be tracked through observations of emission lines.  It may be
expected that jets launched by the magnetocentrifugal mechanism may have a
Keplerian rotation profile, as it will be launched from an extended region
of the disc.

We presented two three dimensional (3D)
magneto-hydrodynamics (MHD) simulations of protostellar jets,
with two different initial magnetic field configurations (OP and BP
configurations; see section~\ref{numericalsection}) in \citet{staff10}.  
In that paper, as in this, we focused on
magnetic field configurations, whose poloidal magnetic field strength at the
disc surface falls off as power laws with disc radius ($r_o$) as $B_p
\propto r_o^{\mu - 1}$ where $ -0.5 \le \mu < 0$ defining the degree of
inclination and decline of the field along the surface of the disc
\citep{jorgensen01}.
The simulation box in
both cases was 60 AU long, and 30 AU wide in the two directions perpendicular
to the jet.  One interesting result was that in the OP
simulation (with the less open initial magnetic field configuration), a
two-component jet structure developed.  That is, a narrow, inner jet
surrounded by, but separated from, a cylindrical shaped outer jet.  No such
outer jet was found in the BP simulation.  However, the outer part of the BP
jet was starting to leave the grid, and it was unclear if this was the
reason for the lack of an outer jet.  This therefore motivated us to
increase the size of the simulation box, and repeat the OP and BP
simulations.  In addition, we also ran two simulations with new initial
magnetic field configurations, $\mu=-0.12$ which is an intermediate
configuration between the OP \citep[$\mu=-0.01$,][]{ouyed197} and BP
\citep[$\mu=-0.25$,][]{bp82} configurations
(see section~\ref{numericalsection} for an explanation of the $\mu$
parameter), and PP \citep[$\mu=-0.5$,][]{pelletier92}
which is even more open than the BP configuration. This way, a more
systematic study of the effect of the initial magnetic field configuration
on the resulting jet can be made. In this paper we present
the results of these four simulations.

In \citet{staff10} we computed synthetic $[OI]~\lambda6300$ emission lines
based on the MHD simulations presented there. Using these emission lines, we 
found good
agreement between the simulations and observations for the jet width,
temperature, density, velocity, and the mass flux in the jets.
We have again been
focusing our effort on computing synthetic emission line maps from our 3D
simulations, and use these to interpret the HST observations of YSO jets.
In this work we show that jets may have a Keplerian rotation profile,
although not in all cases.

Velocity gradients in jets perpendicular to the jet axis can be interpreted
as jet rotation, as \citet{choi11} did for NGC 1333 and \citet{lee08} did
for HH 212.  RW Aur has also shown a velocity gradient perpendicular to the
jet axis \citep{woitas05}, however, \citet{cabrit06} found that this was
opposite to the disc rotation.  \citet{coffey12} re-investigated the RW Aur
jet and found a velocity gradient in a direction opposite of that found by
\citet{woitas05}, but consistent with the disc rotation.  However, later
observations did not show this gradient any more.  While these
velocity gradients may be indicative of jet rotation, the findings by
\citet{coffey12} illustrates that the explanation may not be that simple. In
this paper we investigate the rotation in disc winds, and
offer explanations for these seemingly contradictory observations.

Using the Hubble Space Telescope, nearby protostellar jets in the Taurus
cloud at a distance of 140 pc can be resolved down to 14 AU scale.  This is
very interesting as it allows observations of the region where the jet is
generated and collimated \citep[for a summary, see][]{ray07}.  Also of
importance is that in recent years, increased computer power has allowed 3D
MHD simulations of protostellar jets to be extended to large scales while
still being able to resolve and simulate the source. \citet{anderson06} used
a simulation box extending out to 12 AU, while \citet{moll09} used a
simulation box extending out to 60 AU, a similar length to what we used in
\citet{staff10}. As can be inferred, these latter simulations can cover
several pixels of jets observed by the HST, and that is important as it
allows for direct comparison between the properties of observed jets and
simulated jets. In this work we have extended the simulation box from
\citet{staff10} to 90 AU.

 This paper is organized as follows:   We start by briefly describing our
numerical setup in section 2.  In section 3 we present the tools and the
strategy we use to compute emission lines, PV-diagrams and the visualization
of our simulations.  Results are presented in section 4 with a summary and a
conclusion given in section 5.

\section{Numerical approach}
\label{numericalsection}

We use the ZeusMP code \citep{norman00} to simulate a protostellar jet being
launched from a Keplerian accretion disc.  The simulation setup discussed
here is based on \citet{staff10}, however, the simulation box has been
extended to $3000 r_i$ in the jet direction (from $2000 r_i$ in the previous
paper), and to $\pm 900 r_i$ in the two directions perpendicular to the jet
axis (from $\pm 500 r_i$).  For reasons outlined in \citet{ouyed03} we use a
Cartesian coordinate system ($x_1$, $x_2$, $x_3$). 
In addition we break the quadrantal symmetry by wobbling the disc, as
described in \citet{ouyed03}.  The grid is 1536 zones in the $x_1$ direction
(i.e.  the direction of propagation of the jet), and 500 zones in the $x_2$
and $x_3$ directions (the directions parallel to the plane of the disc). 
This grid is separated into a uniform grid of $200\times100\times100$ zones
located just above the disc and centered around the axis that extends from
-25 to 25 $r_i$ in the $x_2$ and $x_3$ directions, and from 0 to 100 in the
$x_1$ direction.  The remainder of the large grid (200 zones on either side
of the uniform grid in the $x_2$ and $x_3$ directions, and 1336 zones in
extension to the uniform grid in the $x_1$ direction) is ratioed, extending
the large grid from $-900 r_i$ to $900 r_i$ in the $x_2$ and $x_3$
directions and out to $3000 r_i$ in the $x_1$ direction.  Although the
boundaries in our simulation are outflow boundaries, it is important that
very little mass leaves the grid, as this may lead to incorrect results.  We
avoid this by using such a large simulation box.

In the simulations presented in this paper, as well as in many previous
works, the disc is treated as a boundary condition
\citep[e.g.][]{ouyed197,krasnopolsky99,ouyed03,anderson06,staff10,ramsey11}. 
The protostar in our simulations is located at the $x_1=0$ boundary as in
\citet{ouyed03}.  More recently, a few attempts have been made to simulate
self-consistently both the disc and the jet launching
\citep[e.g.][]{zanni07,murphy10,sheikhnezami12}. The fixed physical
conditions in the accretion disc is another key
simplification in our model.  From experience, properly simulating the disc
and reaching scales that we are simulating                               
are beyond the current state-of-the-art simulations.  Some attempts have
been made by various groups at simulating the disc and the jet in 3D but
these simulations do not capture the acceleration, and collimation, and far 
out regions which is our main goal.  In part, this simplification may be
justified by the fact that typically, accretion discs will evolve on longer
timescales than their associated jets.

The density profile of the  hydro-statically stable accretion disc corona  is given by  $\rho=r^{-3/2}$, for $r<1000 r_i$, where 
$r=\sqrt{x_1^2+x_2^2+x_3^2}$.  The density profile of the corona smoothly transits to a uniform density for  $r>1000 r_i$.
This configuration remains stable and cures  numerical issues induced by extremely low density in  the coarse grid. This
does not affect the outcome of the simulations. 
The density profile in the disc matches that of the overlaying corona except for 
 much higher density in the disc (i.e. a density jump of a factor 100 between the disc and the corona)
 imposed by pressure balance.  The corona and the accretion disc (a fixed
boundary in our simulations) are threaded by an initially purely poloidal,
force free and current free, magnetic field.  
Most accretion disc models used assume power law distributions of
density and temperature, so it is natural to assume power law structure for
the fields, i.e. $B_p\propto r_0^{\mu-1}$.

The velocity profile in the disc is Keplerian, $v_{\phi}\propto  
1/\sqrt{r}$.    At the inner and outed edges of the disc, the velocity profile is smoothed to
avoid a cusp as discussed in \citet{ouyed03} (see Appendix A1
 and   simulation E in that paper).  The outer edge of the disc was set to $r_{\rm out}= 80 r_i$.  However,
 the outer edge has less impact on our results since in most of our simulations
 the jets are launched from close the innermost regions of the disc. So as long
 as $r_{\rm out} >> r_{\rm i}$  increasing $r_{\rm out}$  does not make much
 of a difference. 
 
 For our fiducial values (i.e. setting the mass of the protostar  to $1 M_\odot$),
  we estimate $r_{\rm in}\simeq 0.03$  AU \citep{ouyed197} assuming that the
stellar magnetosphere truncates the disc roughly at this radius.
  For a comparison,  recent studies of the inner circumstellar disc of S CrA
N (a T Tauri star) with the VLTI suggest that $r_{\rm in}\sim 0.11$ AU
 for a corresponding $\sim 1.5 M_\odot$ protostar \citep{vural12}.
Hence, the simulations are run on a grid extending out to
90 AU along the jet axis, and 27 AU on either side of the jet axis 
\citep[compared to 15 AU on either side of the jet axis and 60 AU long in][]{staff10}.

\subsection{Parameters in the model}

As listed in \citet{staff10} \citep[see also][]{ouyed03} the parameters of
the model can be divided into disc parameters and those defining the corona. 
The parameters related to the disc are: (i) The injection speed $v_{\rm
inj}$ which is set to a tiny fraction of the Keplerian speed ($v_K$) so that
$v_{x_1}=v_{\rm inj} v_\phi$ with $v_{\rm inj}=0.003 v_K$ in all of our
simulations, which makes it subsonic.
  The injection speed defines the  mass loading. 
As discussed in \citet{ouyed99}, $v_{\rm inj}$
has an upper bound set by the fact that we want the sonic point to be
resolved in the simulations. Numerically it is hard to lower $v_{\rm inj}$
much, as mass is being depleted from the region just above the disc faster
than it is being resupplied from the disc. Such low density regions can
easily cause problems in numerical simulations like these;
(ii)  Pressure balance between the corona and the underlying disc fixes the
 density jump between the two. This  defines  another parameter  $\eta_i=100$ which is the ratio
 between the disc density and the density in the overlaying corona;.
 The remaining parameters define the corona: (i) the corona is kept in hydrostatic
 balance by thermal pressure which defines a parameter $\delta = 5/2$;
 (ii) The ratio of gas pressure to magnetic pressure in the corona at $r_{\rm i}$ (i.e at the inner disc 
boundary) was set to equipartition value; i.e. $\beta=1$.

The magnetic field configuration (given by $\mu$ with $ B_p \propto
r_o^{\mu - 1}$ along the disc's surface) is the one parameter which is
varied.  In order to investigate the role of the initial
magnetic field geometry in launching and collimating jets, we analyze 
results from four configurations: The
Ouyed-Pudritz (OP with $\mu = - 0.01$), the Blandford-Payne (BP with $\mu =
- 0.25$), the Pelletier-Pudritz (PP with $\mu = - 0.5$) a fourth
configuration between the OP and BP one with $\mu = -0.12$.

\section{Visualization and diagnostic tools}

\begin{figure*}
\includegraphics[width=1\textwidth]{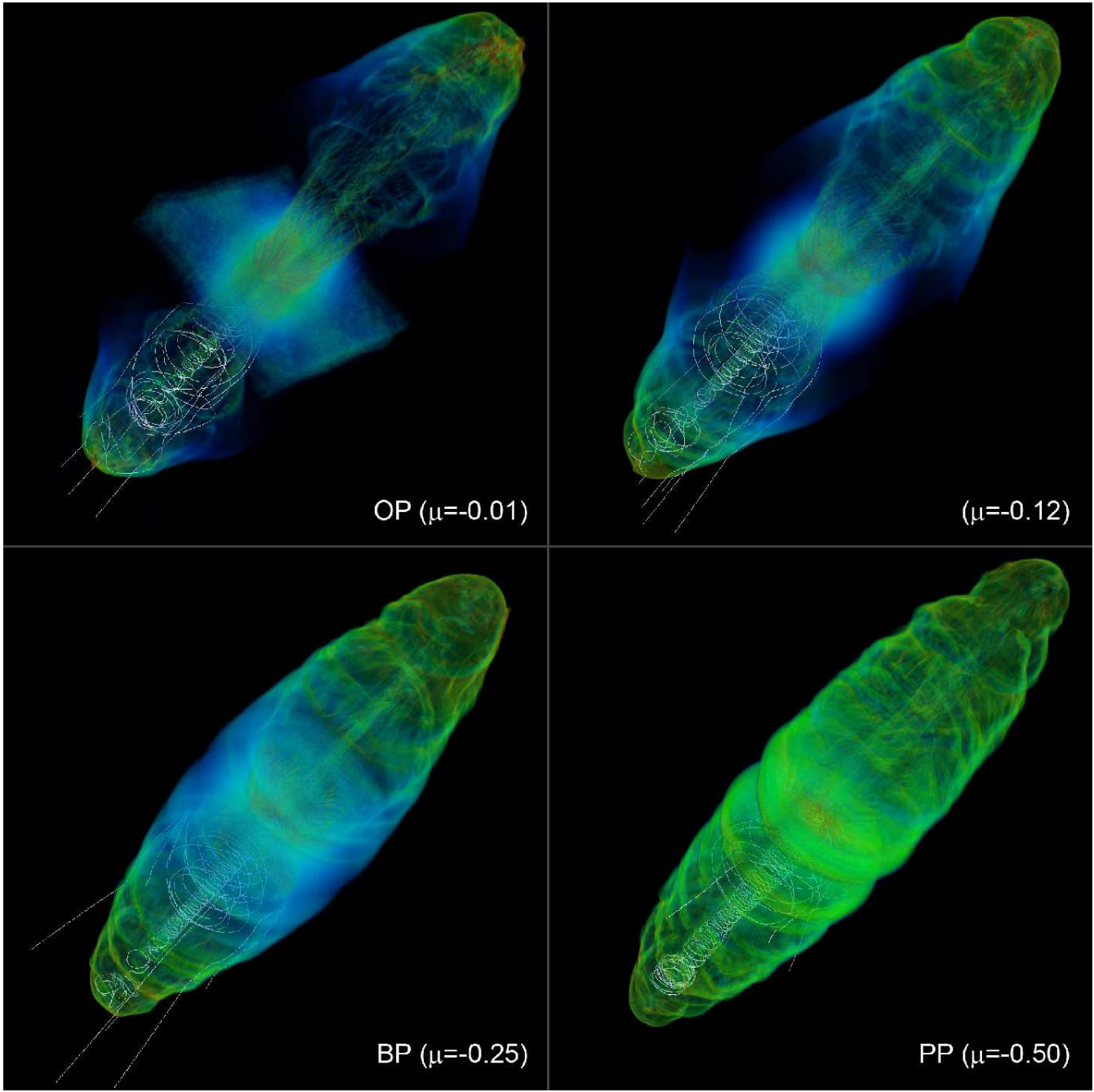}
\caption{3D figures of the final jet structure in the four simulations. The
jet is only simulated in one hemisphere. To visualize it, it is mirrored in 
the disc plane to make it appear bipolar.  The field lines are only drawn in one
hemisphere in each figure, so that the density structure can be best
visualized. See \texttt{http://quarknova.ucalgary.ca/MHD.html} for the
relevant animations. }
\label{visitfigures}
\end{figure*}

\begin{figure*}
\includegraphics[width=\textwidth]{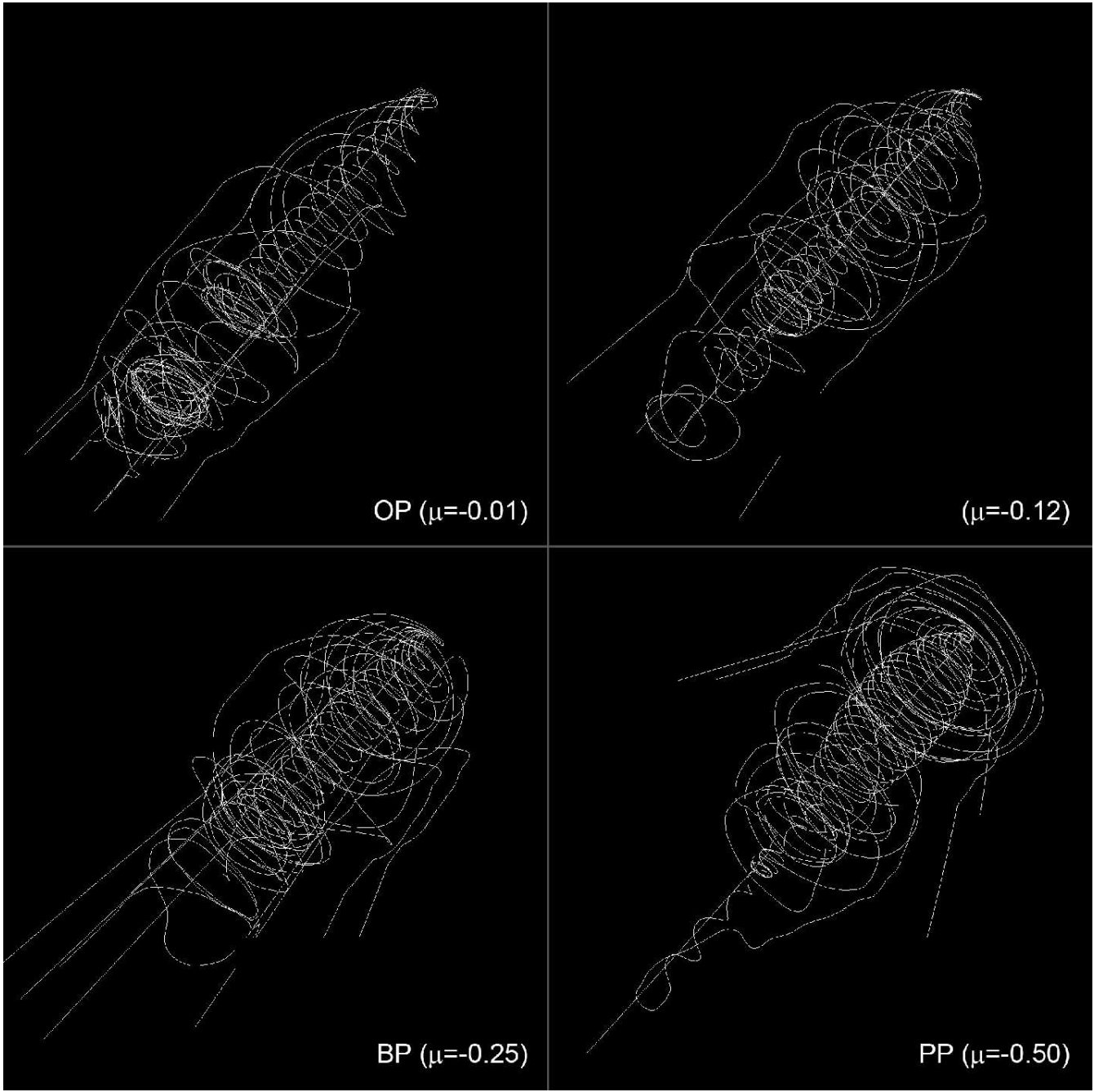}
\caption{The magnetic fieldlines for each jet (only one hemisphere is shown). 
The disc and the protostar are in the
upper right corner of each figure. The PP jet shows a very strong
collimation halfway through the jet (which is also visible in
Fig.~\ref{visitfigures}). The foot-points for the field lines are
chosen slightly differently than in Fig.~\ref{visitfigures}, which is why
the fieldlines look slightly different. However, the general features remain in 
both figures.}
\label{fieldlines}
\end{figure*}

\begin{figure*}
\includegraphics[width=\textwidth]{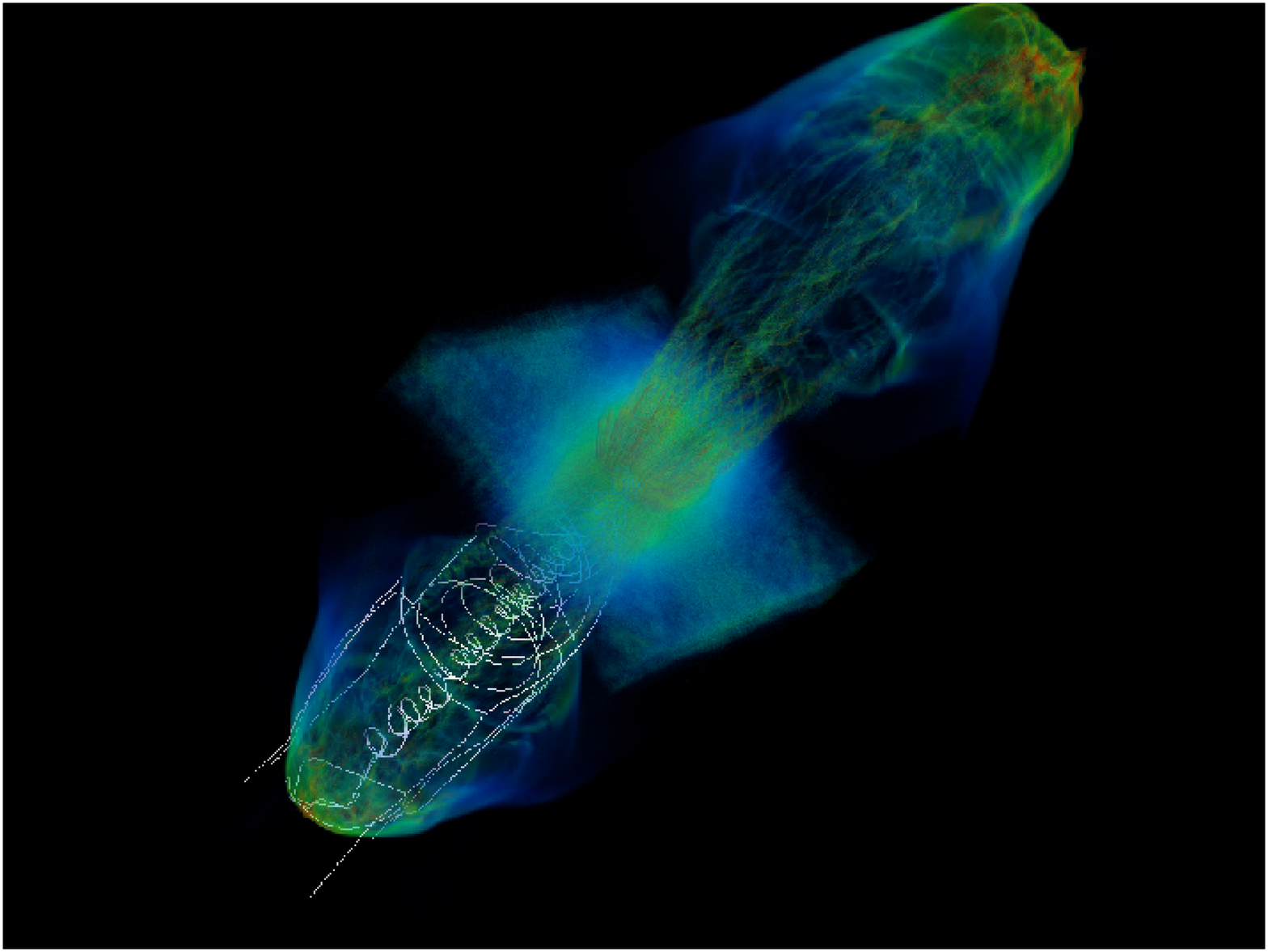}
\caption{A larger version of the OP jet from Fig.~\ref{visitfigures} at the
end of the simulation, shown 
for better clarity. The 2 component jet structure (i.e. a narrow inner jet
surrounded by a cylindrical shaped outer jet) is clearly visible, as are
the twisted magnetic field, especially around the inner jet. The bow shock 
at the front of the jet is also clearly visible. The protostar and the disc is hidden in the middle.
As in Fig.~\ref{visitfigures} we only draw the magnetic field lines in one
hemisphere to better illustrate the details of the density structure.}
\label{OPvisitfigure}
\end{figure*}

We visualize the simulations using two softwares namely Visit (used to
generate Figures~\ref{visitfigures}, \ref{fieldlines}, and \ref{OPvisitfigure}
in this paper; see
https://wci.llnl.gov/codes/visit/) and the software SHAPE \citep{steffen11}
for the other figures.  SHAPE is capable of visualizing HDF data in a
variety of forbidden emission lines making it an ideal tool for generating
observation comparable images and PV diagrams.

SHAPE implements a ray-casting radiative transfer algorithm to calculate the 
emission reaching the observer.  For each pixel in the final image, a ray 
is cast from the back of the 3D grid to the front (observer).  As the ray 
passes through the grid, emission coefficients are calculated according to 
the density, temperature and species contained within the grid cell.  For 
the current simulations, we assume a 2-level system under optically thin 
conditions in LTE and calculate the emission coefficients for [ClIV]
$74500~{\rm \AA}$, [SII] 
$6730~ {\rm \AA}$, and MgII $2796~ {\rm \AA}$.  The line data for these 
transitions were obtained from the Chianti spectral line database 
\citep{dere97}. The critical densities for these lines are $\sim 1$ cm$^{-3}$,
 $\sim 10^{4}$ cm$^{-3}$, and $\sim 10^{15}~$cm$^{-3}$ for the [ClIV],
[SII], and MgII transitions respectively.  Typical electron densities in our
simulations are around $10^4~{\rm cm^{-3}}$ (assuming an electron fraction
of 0.1). 
Therefore our three lines represent transitions with critical densities well
below, equal to, and well above the jet densities.  For densities much less
than critical, the emission depends on $n^2$, and for densities much higher
than critical, it depends on $n$ \citep[see eg.][]{osterbrock06}.  Our chosen lines therefore represent both
extreme cases and the transition from one regime to the other.

 The dimensionless temperature of the gas used in the calculation of the
emission coefficients can be found by using a polytropic EOS
\citep[$P\propto\rho^\gamma$, $\gamma=5/3$; see][]{staff10}: $T=\rho^{\gamma-1}$. 
If the Alfv{\'e}n Mach number ($M_{\rm A}$) is greater than 0.3 we assume
the gas to be shocked and the temperature is then
  found by using $T=\rho^{\gamma-1}\frac{M_A^2}{\beta}\frac{\gamma-1}{2}$
where $\beta=P_g/P_B$ is the plasma $\beta$ 
\citep[for details, see][]{ouyed93}.

With these tools we can present our  simulations  as seen in [ClIV], [SII],
 and MgII.  The [SII] and MgII lines are known to be strong, leading to
 higher signal to noise ratio when observing real jets
 \citep[e.g.][]{bacciotti00,coffey12}.
Finally, SHAPE allows us to easily generate Position-Velocity diagrams which
as we show later turn out to be crucial when comparing the different
simulations.

Our use of a polytropic equation-of-state is a limitation. Nevertheless,
this ansatz (which effectively by-passes the energy equation) corresponds to
situation where the combined effects of cooling and heating simulates a
tendency towards a state of local constant entropy.  This also tremendously
simplifies the numerical setup.  Except in cases where the
 shocks are dynamically important \citep{ouyed03}, this simplification
allows us to test the code against analytical solutions \citep[e.g.][]{bp82}. 

We compute the radiation in the post-processing step (using Shape), assuming
heating from shocks which may appear inconsistent with the previous
statement, while at the same time assuming that neither the radiation nor
the cooling plays a dynamically important part in the simulations.  These
are simplifications that are necessary at this point, but should be
addressed in future calculations.  Likewise, we assume a constant  
ionization fraction, while in reality it may well vary throughout the jet
and also change over time, for instance in shocks.

\section{Results}

Figure~\ref{visitfigures} shows the final density and magnetic field
structure of the jets with different magnetic field structure on the disc
(different $\mu$).  The protostar and the disc is located in the middle of
the figure.  For all the different magnetic field configurations on the
disc, we find a narrow jet extending out along the axis.  Close to the disc
and the protostar it is relatively straight, but farther away it starts to
twist up in a corkscrew-like fashion (see also
section~\ref{rotationsection}).  In the front of the jet, a bow shock is
clearly visible.  More negative $\mu$ leads to a higher Mach number, and
therefore the bow shock is bent more strongly backwards in those cases. 
Extending backwards from the bow shock is a cocoon of material that have
been processed by the bow shock.
In the BP and PP simulation, this cocoon stays on the grid
throughout the entire simulation, while in the OP and $\mu=-0.12$
simulations part of this material is pushed off the grid.

Figure~\ref{fieldlines} shows the field lines only, and only in one
hemisphere, for better clarity. Independently of the magnetic field on the
disc, the field lines
wrap up tightly around the narrow jet, indicating that the hoop stress is
collimating the jet. 
In the PP jet we even find tighter collimation of the
field lines further out in the jet, indicating increased collimation of the
jet at larger distances. For all the magnetic field configurations,
there is also a ``spine'' in the
middle of the jet of mostly poloidal field that helps stabilize the jet
\citep[for more about jet stability, see][]{ouyed03}.
This mostly poloidal field originates from the protostar (not the disc),
explaining why it is not being wrapped up (the foot-point is not orbiting). 
The most tightly wound-up field
originates in the very inner part of the disc. For all the configurations,
some field lines do not wind-up but are seen to be mostly straight. These
field lines originates just a few AU into the disc, illustrating that the
jet originates from the very innermost one or two AU of the disc.

\subsection{The two-component jet}

As was pointed out in \citet{staff10}, the
OP jet is a two-component jet, consisting of a narrow, inner jet, and
surrounded by a cylindrical shaped outer jet.  The BP simulation in
\citet{staff10} did not show any such outer jet, and it was questioned
whether this was related to the fact that mass had started flowing off the
side of the grid in that simulation.  With the bigger grid, we here again
find that the OP jet has a two-component structure like found before
(collimated by the less tightly twisted field lines anchored farther out in
the disc), and we
find only one clear component in the BP jet, although there is a very faint outer
jet surrounding the inner part (close to the $x_1=0$ boundary) of the BP
jet.  The $\mu=-0.12$ configuration, which is intermediate between OP
($\mu=-0.01$) and BP ($\mu=-0.25$), also results in a clear outer jet
component, at least for the first half of the jet.  There is no outer jet
seen in the PP simulation. This can be seen in the density structure in
Fig.~\ref{visitfigures}. 

We show a larger
version of the OP jet in Fig.~\ref{OPvisitfigure} to better illustrate the
features of that jet. Outside the inner jet, other field lines with
foot-points farther out in the disc, are seen to wrap up, although less
tightly. We  find that field lines with footpoint within $\sim15 r_i$
($\sim0.45$ AU) in the OP jet twist up thightly around the inner jet. Field
lines with footpoints outside of this are less thightly wound up and leads
to the outer jet. Beyond $\sim50 r_i$ ($\sim 1.5$ AU), field lines do not twist
up much but are mostly straight and we do not expect these to contribute
much to the jet. In contrast to this, field lines with footpoint inside of
$\sim20 r_i$ ($\sim 0.6$ AU) in the PP simulation twist up thightly around
the jet. Footpoints farther out twist up with a much larger radius,
but do not propagate much forward along the jet (see Fig.~\ref{fieldlines}) and will 
therefore not contribute
to any jet. This is why there is no outer jet in the PP simulation.

In the $\mu=-0.12$ simulation, the bow shock is also bent more sharply backwards
than in the OP simulation, and the outer jet is therefore terminated earlier
upon interacting with the cocoon.  Thus, it seems that the shape of the bow
shock and the resulting back-flowing current affect the
evolution of the second jet component.  

\begin{figure*}
\includegraphics[width=\textwidth]{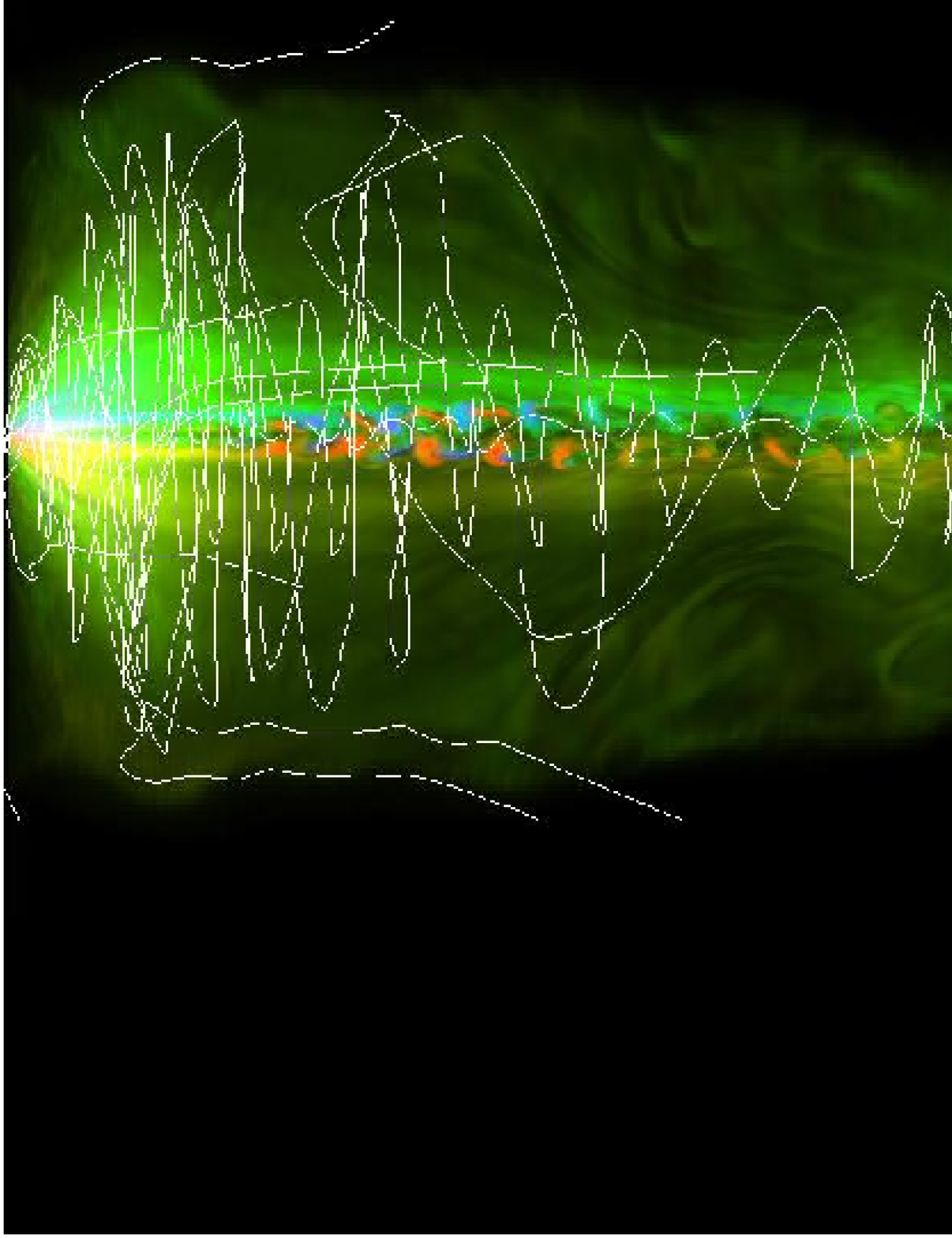}
\caption{Magnetic field lines and toroidal velocity for the PP jet. The disc
and protostar are on the left boundary. The brighter colors illustrates
higher intensity radiation. The field lines
show an increased collimation about halfway through the jet. At the same
place, the high intensity part of the jet 
narrows, showing that in the PP jet the collimation increases
further out in the jet.}
\label{pp_vz_lines}
\end{figure*}

\begin{figure*}
\includegraphics[width=\textwidth]{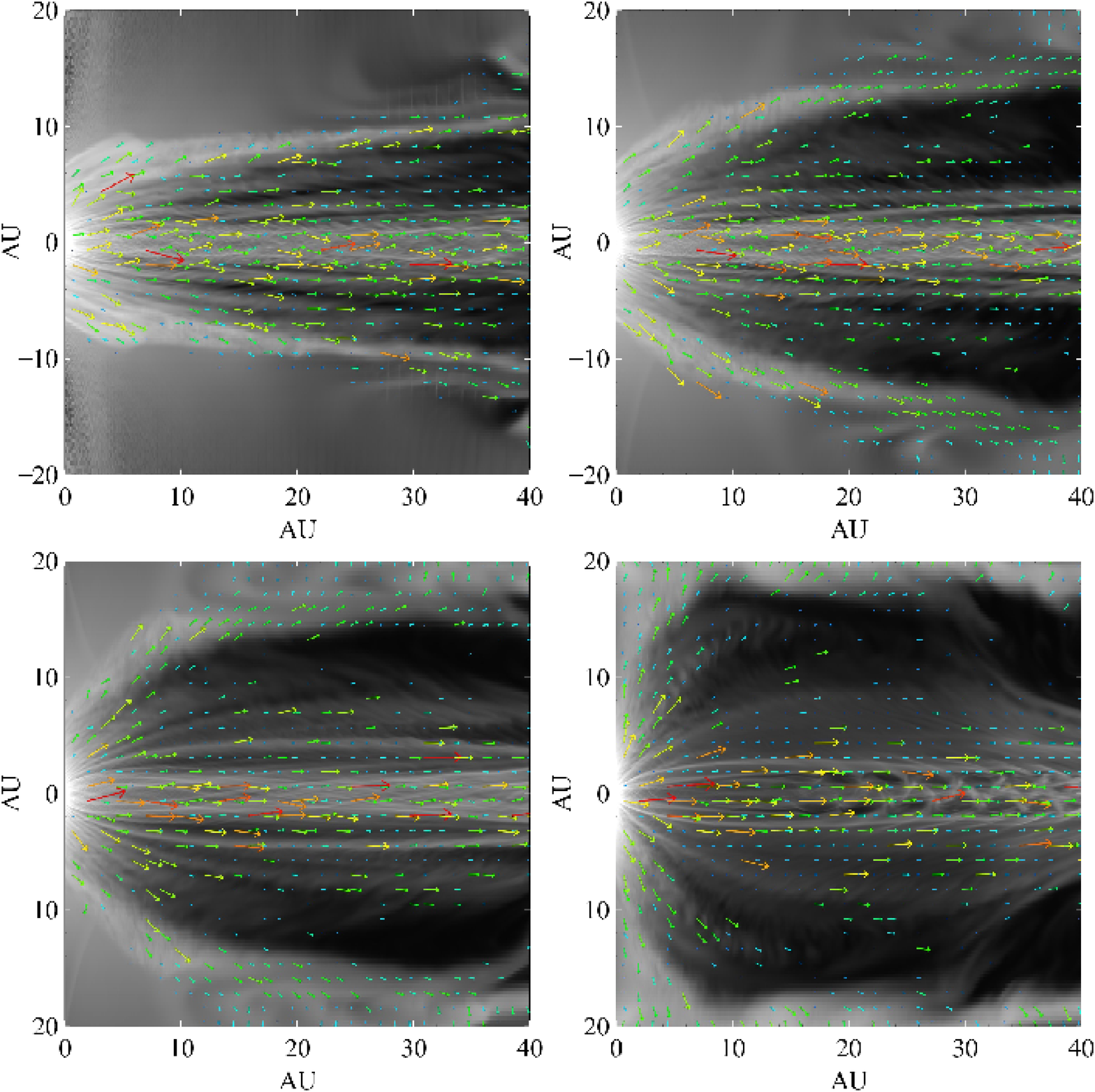}
\caption{Momentum-density vectors and density shown in a plane along the jet
axis. This particular view is focused on a region closer to the disc. The
PP jet shows more momentum flowing sideways away from the axis
compared to the OP jet in particular. With the weaker magnetic field far
from the axis in the PP case, this can not easily be directed forward,
explaining the lack of an outer jet in the PP simulation. Higher momentum
vectors are read, lower momentum vectors are blue. The vectors do not have
the same normalization for the different simulations.}

\label{momentum}
\end{figure*}

The recollimation effect of the PP jet is also reflected in the toroidal
velocity, as shown in Fig.~\ref{pp_vz_lines}. The rotating jet narrows as
the twisting field lines also narrow, showing that the jet itself is more
collimated farther away from the disc.
We show in Fig.~\ref{momentum} the mass density and the momentum density for
each simulation in a slice along the jet axis, zoomed in on a region close
to the disc. In the OP jet, 
the momentum-density vectors point mostly forward. For the
initially more open magnetic field configurations, the
momentum-density vectors points less forward, and more away from the axis.
In the PP jet, material is flowing almost perpendicular away from the axis 15-20 AU
from the axis, close to the disc boundary. 

\begin{figure*}
\includegraphics[width=0.95\textwidth]{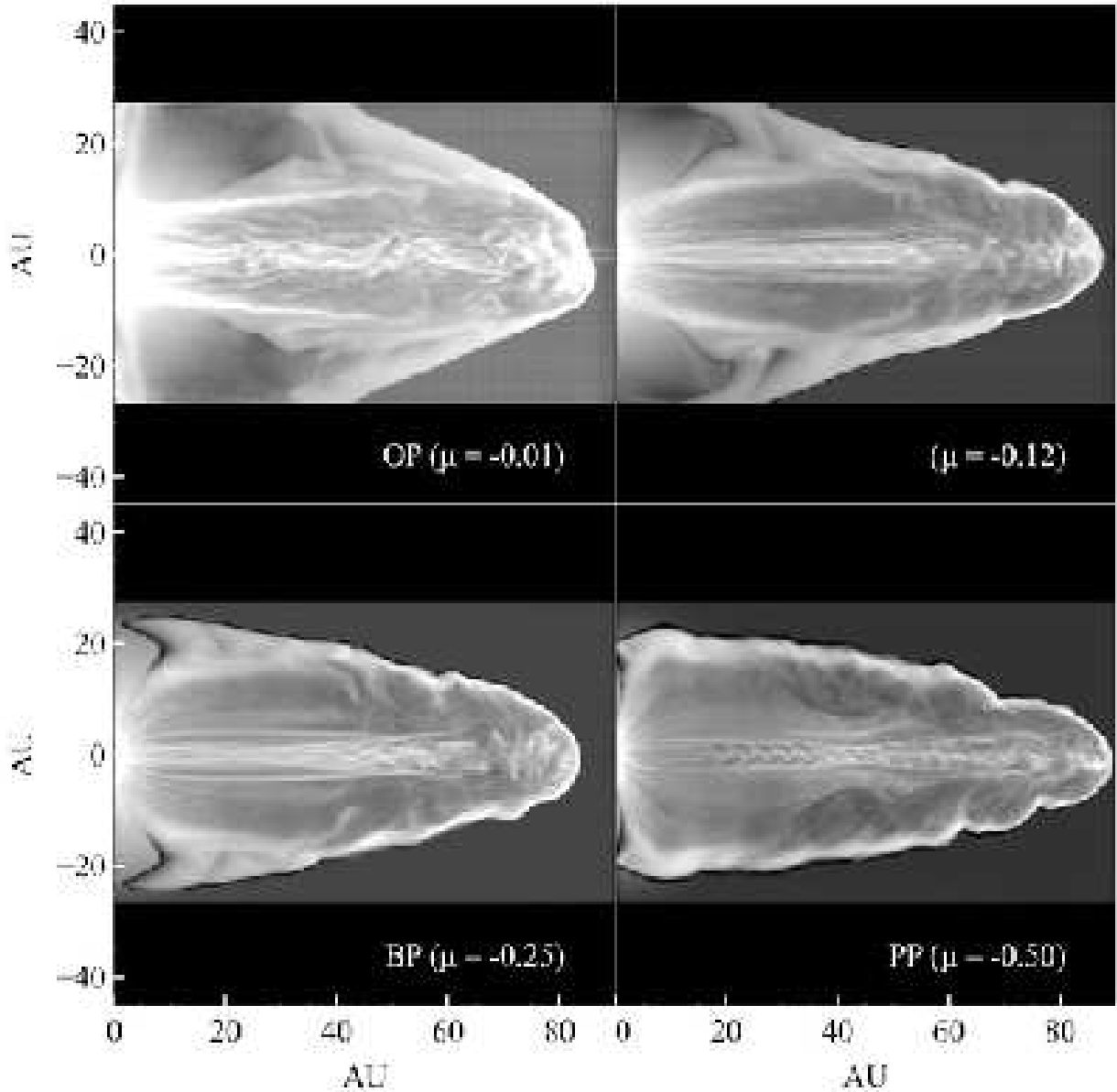}
\caption{The OP, $\mu=-0.12$, BP and PP jets seen in [ClIV] $74500~ {\rm
\AA}$.  The x-axis range is from 0 - 90 AU and the y-axis from -45 to +45 AU 
for each image. The disc and the protostar are on the left boundary in these
images. See \texttt{http://quarknova.ucalgary.ca/MHD.html} for animation.
The normalization factor is different from that in Figs.~\ref{imageSII}
and~\ref{imageMgII}, and the jet would appear much dimmer in this [ClIV]
line than in the [SII] or MgII line.}
\label{imageClIV}
\end{figure*}

\begin{figure*}
\includegraphics[width=0.95\textwidth]{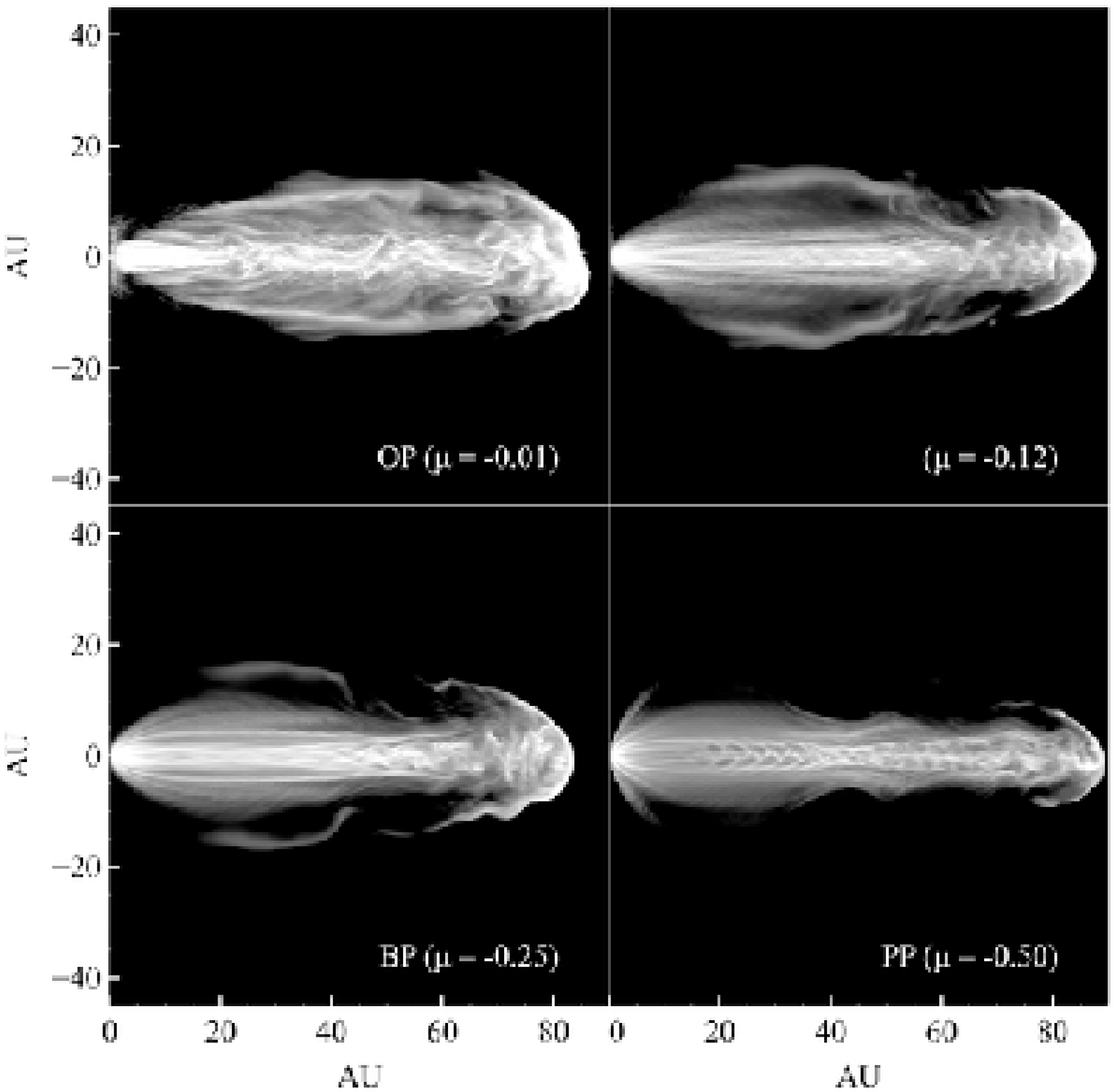}
\caption{The OP, $\mu=-0.12$, BP and PP jets seen in [SII] $6730~ {\rm
\AA}$.  The x-axis range is from 0 - 90 AU and the y-axis from -45 to +45 AU 
for each image. The disc and the protostar are on the left boundary in these
images. See \texttt{http://quarknova.ucalgary.ca/MHD.html} for animation.
The normalization factor is different from that in Figs.~\ref{imageClIV}     
and~\ref{imageMgII}, and the jet would appear much brighter in this [SII]
line than in the [ClIV] but much dimmer than in the MgII line.}
\label{imageSII}
\end{figure*}

\begin{figure*}
\includegraphics[width=0.95\textwidth]{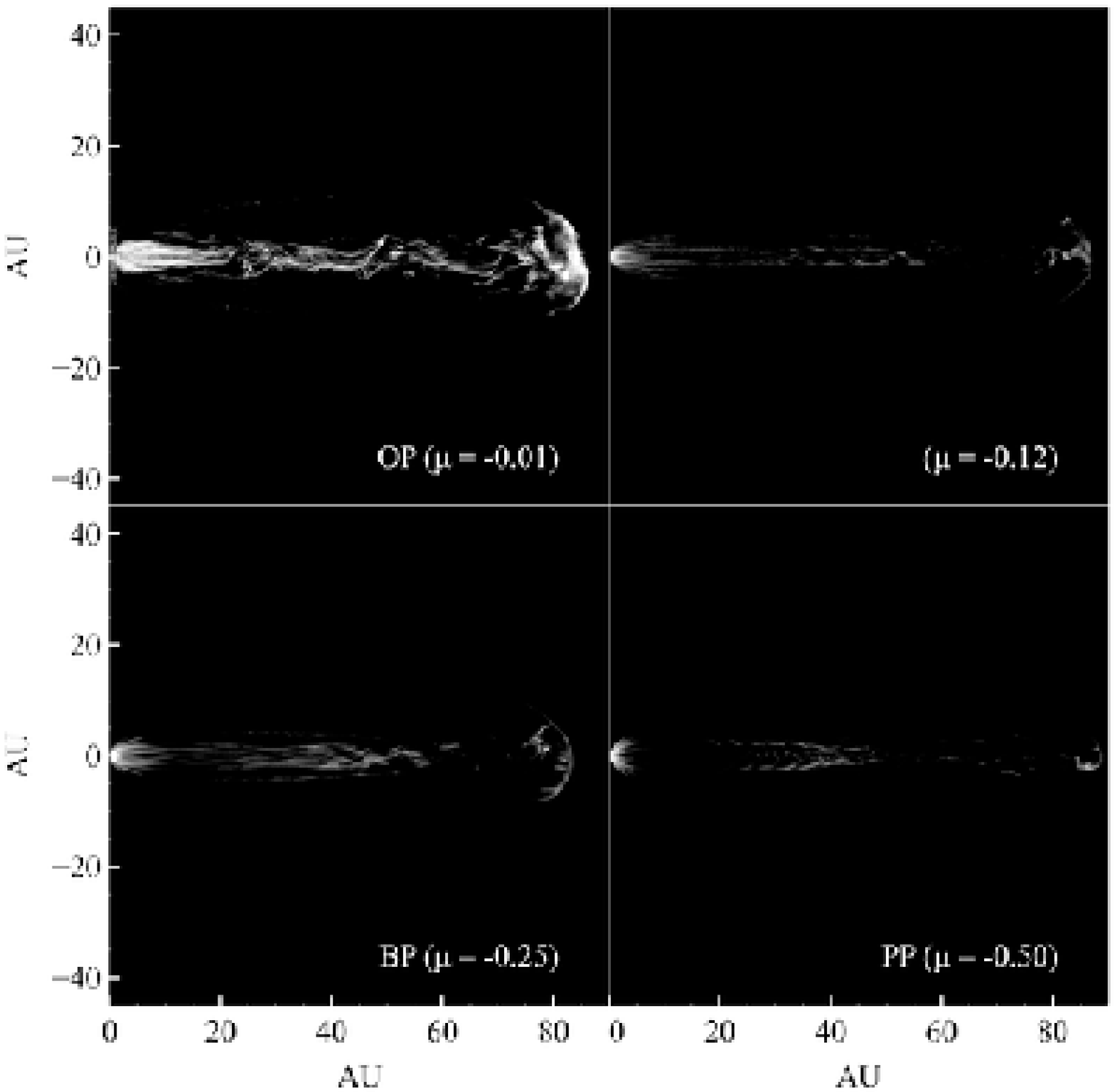}
\caption{The OP, $\mu=-0.12$, BP and PP jets seen in MgII $2796~ {\rm \AA}$.  The
x-axis range is from 0 - 90 AU and the y-axis from -45 to +45 AU for each
image. The disc and the protostar are on the left boundary in these
images. See \texttt{http://quarknova.ucalgary.ca/MHD.html} for animation.
The normalization factor is different from that in Figs.~\ref{imageClIV}     
and~\ref{imageSII}, and the jet would appear much brighter in this MgII
line than in the [ClIV] and [SII] line.}
\label{imageMgII}
\end{figure*}

\begin{figure*}
\includegraphics[width=.95\textwidth]{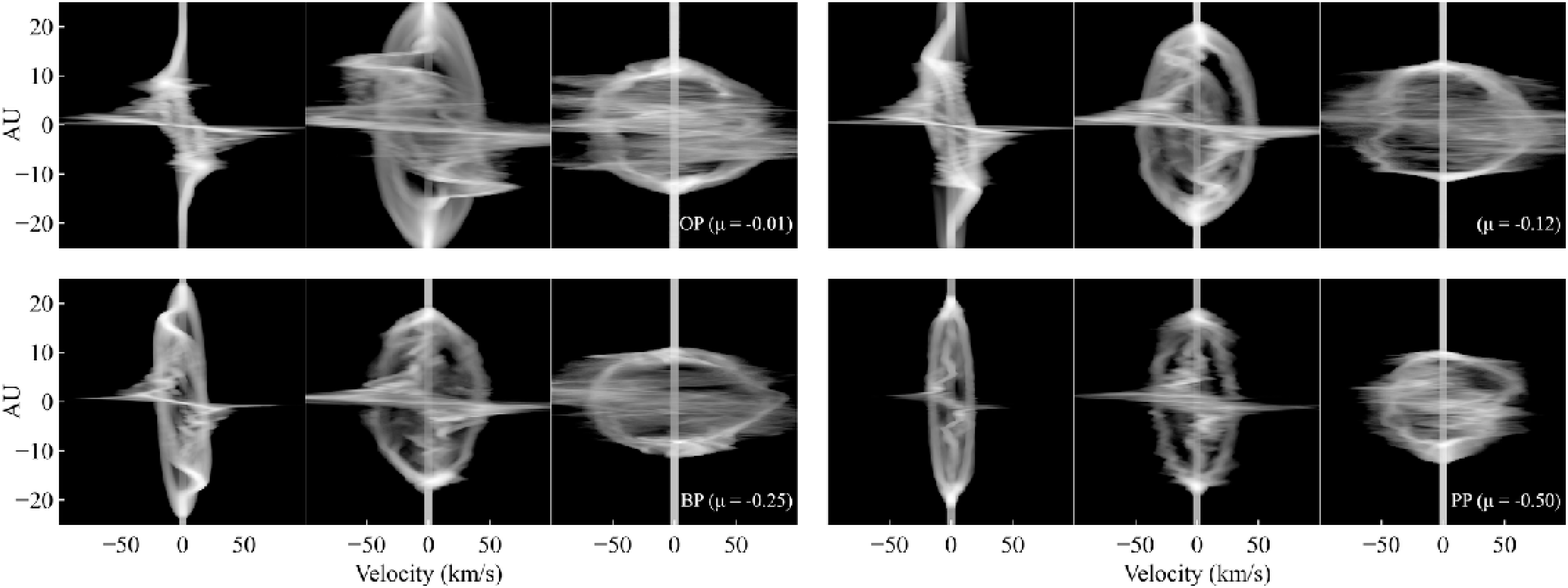}
\caption{PV diagrams for the OP, $\mu=-0.12$, BP and PP jets as seen in
 [ClIV].  The slit locations are perpendicular to the jet at 15, 45, and 
75 AU from the disc represented in the left, middle and right panels 
respectively.}
\label{pvClIV}
\end{figure*}

\begin{figure*}
\includegraphics[width=.95\textwidth]{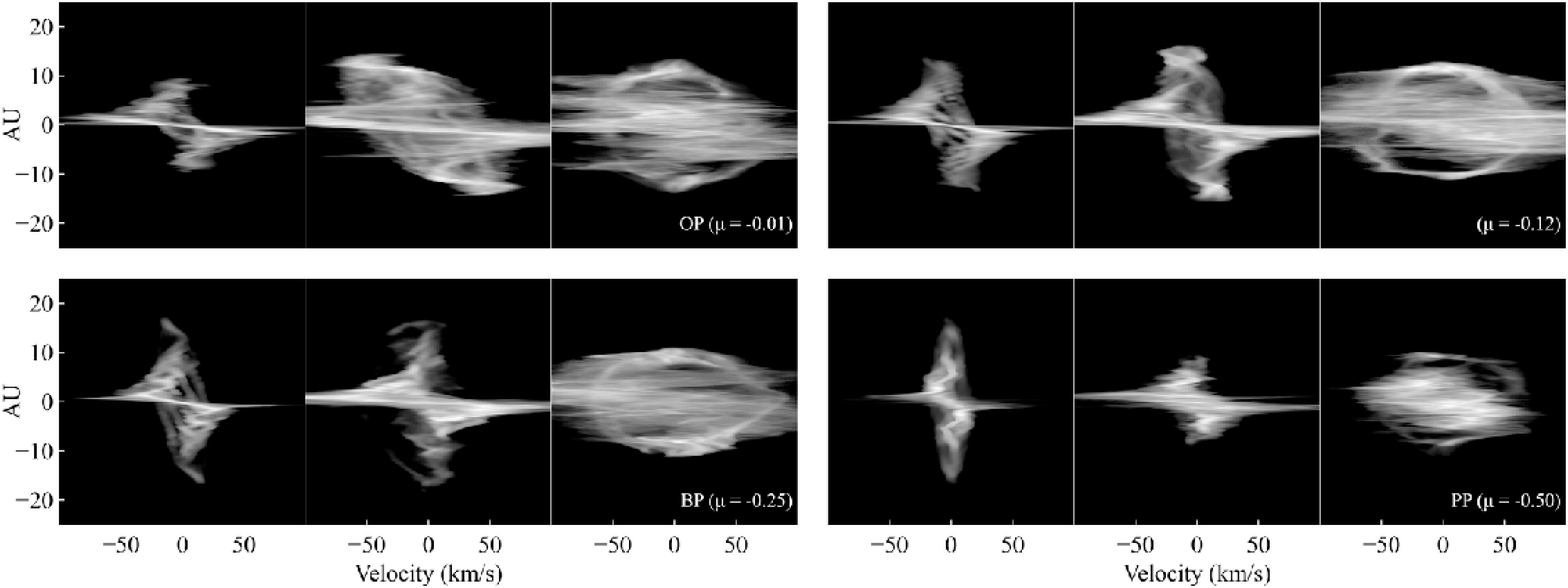}
\caption{PV diagrams for the OP, $\mu=-0.12$, BP and PP jets as seen in [SII].  The slit locations are perpendicular to the jet at 15, 45, and 75 AU from the disc represented in the left, middle and right panels respectively.}
\label{pvSII}
\end{figure*}

\begin{figure*}
\includegraphics[width=.95\textwidth]{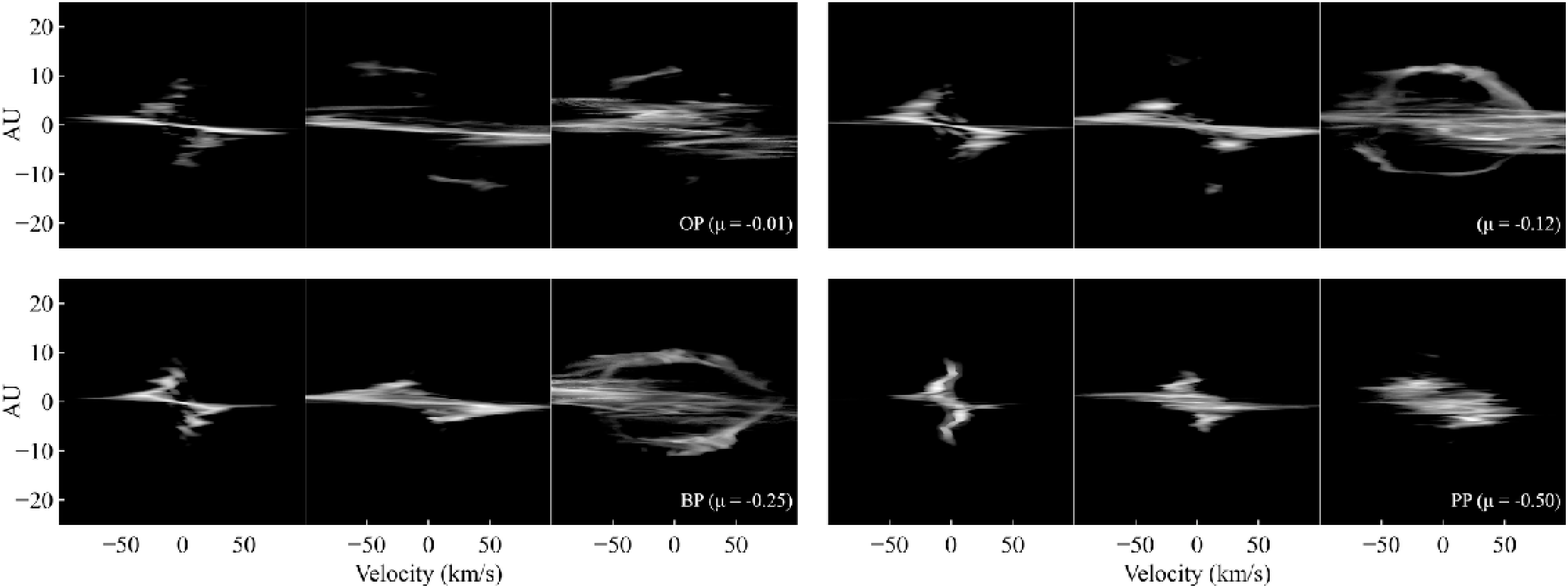}
\caption{PV diagrams for the OP, $\mu=-0.12$, BP and PP jets as seen in MgII.  The slit locations are perpendicular to the jet at 15, 45, and 75 AU from the disc represented in the left, middle and right panels respectively.}
\label{pvMgII}
\end{figure*}

In Figs.~\ref{imageClIV}, \ref{imageSII}, and \ref{imageMgII} we show the
different simulated jets as they appear in [ClIV] $74500~{\rm \AA}$, [SII]
$6730~ {\rm \AA}$ and MgII $2796~ {\rm \AA}$ respectively.  In
Figs.~\ref{pvClIV}, \ref{pvSII}, and \ref{pvMgII} we present
position-velocity (PV) diagrams in the same lines taken perpendicular to the
jet at 15, 45, and 75 AU from the disc.  
Due to their vastly different critical densities, the jet appears quite
different in each line.
In the MgII line image, the core of the jet is much brighter than the
surrounding cocoon, and it completely dominates the picture.  For the [SII]
line, parts of the core of the jet is above the quenching density, while the
outer parts are below.  As a consequence, the core of the jet is not that
bright compared to the surrounding cocoon, and hence this can be seen as
well.  For the [ClIV], the quenching density is below the density in most of
the jet.  We note that the emission in Figs.~\ref{imageClIV},
\ref{imageSII}, and \ref{imageMgII} is not normalized to the same factor. 
As a consequence the jet would appear orders of magnitude dimmer in [ClIV]
than in [SII] or MgII.

The emission from the inner, dense part of the jet is much enhanced 
in the MgII line (sinse this is below the critical density), and the
outer jet is not visible at all.
The PV diagram found using this line is quite messy reflecting the structure
of the high densities in the inner jet. 
The outer jet shows up very clearly in the [ClIV] line PV diagram, 
especially for the
OP jet at 45 AU, as a strong increase in the velocity around 10 AU. The
outer jet is also visible in the $\mu=-0.12$ jet at both 15 and 45 AU, and
also in the BP jet at 15 AU. It might be that given enough time an outer jet
will set up as the surrounding cocoon drifts outwards, if enough mass can be
channelled that way. The outer jet is also visible in the [SII] images.
In \citet{staff10}
we identified the outer jet in the OP simulation with a strong increase in
toroidal velocity.  We still see this for the OP and $\mu=-0.12$ jets, but
for the BP jet there is only a weak increase in the toroidal velocity. The
cocoon does not appear to interfere at this point, and we therefore conclude
that no strong outer jet develops in the BP configuration.

The jet width can be read off from the line maps and the PV diagram.  We
find it by looking at the emission line maps (Figs.~\ref{imageClIV},
\ref{imageSII}, and \ref{imageMgII}). The [SII] line gives the best estimate
for the true jet width, as it picks up most of the jet but not the cocoon. 
The [ClIV] sees the entire cocoon surrounding the
jet. On the scale shown here, the MgII line does not see either the outer
jet nor the cocoon for either simulation. We emphasize that this is simply
because of how we separate the contours in our plots. Since the high density
inner jet is much brighter than the outer jet, this is where the contours
are for the MgII line. We also emphasize that in the MgII line, the inner
jet is much brighter than any part of the [SII] or [ClIV] line maps.
We note that the simulated jet
is still very young, and in real jets this cocoon would likely have been
pushed sufficiently far away that it would not confuse the observations.

Ignoring the effects of the cocoon, we find by looking at the line maps of
the configurations with the more open disc magnetic field (ie. the more rapidly declining magnetic field distribution on the
disc, or more negative
$\mu$) that the jets appear most collimated.    This is in part
because of the lack of an outer jet.  The PP jet opens up close to the disc,
but narrows around 35 AU (as discussed earlier).  A somewhat similar effect
can be seen in the BP jet, although it is a much smaller narrowing than in
the PP jet.  The inner jet in $\mu=-0.12$ and the OP jet does not show such
narrowing, rather it appears to continue widening with a somewhat constant
opening angle throughout the length of the jet.  We find, by looking at the
[SII] line emission maps, that the inner OP jet has an opening angle of about 7 degrees.  Again, from
the [SII] line emission map (Fig.~\ref{imageSII}), we find the $\mu=-0.12$ jet  to have an opening angle
of about 5.6 degrees.  
The outer jet appears to re-collimate (around 35 AU for BP, 50 AU for
$\mu=-0.12$, and around 60 AU for the OP jet), but it is unclear if this is
due to the interaction with the cocoon.  These results agree with
\citet{staff10}, where the BP (more negative $\mu$) jet was found to be
narrower than the OP jet.

\begin{figure}
\includegraphics[width=0.5\textwidth]{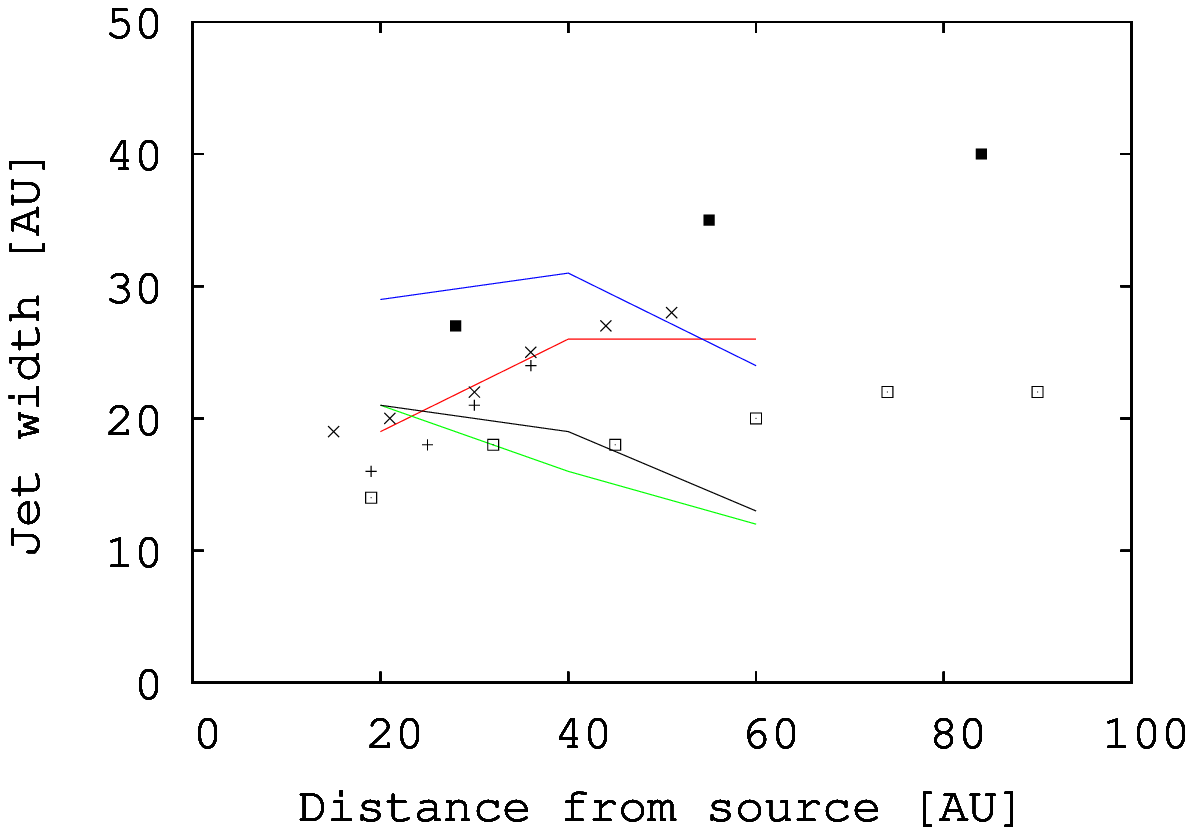}
\caption{The width of the simulated jets as read off the [SII] figures. 
{\it Red:} OP, {\it blue:}
$\mu=-0.12$, {\it green:} BP, {\it black:} PP simulations. The width was
measured at 20, 40, and 60 AU, and these points simply connected with a
line. In addition, we have indicated the width of the DG Tau (filled
squares), RW Aur (open squares), UZ Tau (pluses), and HN Tau (x-es), as read
off from Fig. 2 in \citet{ray07}. }
\label{jetwidth}
\end{figure}

In Fig.~\ref{jetwidth} we plot the width of the simulated jets, and compare
it with the observed width for DG Tau, UZ Tau, HN Tau, and RW Aur. The width
of the simulated jets were found from the [SII] line maps
(Fig.~\ref{imageSII}). As can be seen in Fig.~\ref{imageSII}, the outer jet
in the $\mu=-0.12$ and the BP jets end before the head, making the width
decrease for these two jets at 60 AU. This is likely an artifact of the jet
being so short, had it been much longer, we expect this drop to also occur
much farther out, always just behind the head of the jet. It is interesting
that the width of the HN Tau jet agrees very well with the OP jet. We also
note that the RW Aur jet remains reasonably narrow, making it agree more
with the most negative $\mu$ simulations (BP and PP). 

\citet{hartigan07} found that the HH 30 jet has a constant half opening 
angle of $2.6^\circ$. We have found that for $\mu>-0.25$, the inner jet opening
angle increases with increasing $\mu$. We have also found that the outer jet
becomes weaker for more negative $\mu$. Extrapolating our results, we
conclude that the HH 30 jet has $-0.25<\mu<-0.12$, and likely with $\mu$
closer to $-0.25$ than to $-0.12$. In that case it is likely that the outer
jet is very weak, and it may not be observed.

The jets seen in MgII in Fig.~\ref{imageMgII} appear quite broken up, with
lots of internal structure.  These are not related to density fluctuations
in the jet, but are rather caused by shocks making parts of the jet hotter
than other.  The jets appear more uniform when seen in SII or ClIV, this is
simply a consequence of how we separate the contours, as explained before.
\citet{hartigan07} found knots in the HH 30 jet related to 
higher excitation but without a density enhancement. 
These high excitation knots form on a 100 AU scale, and could potentially be
related to the high excitation knots found in the HH 30 jet.  

From the PV diagrams, we see that the maximum velocity perpendicular to the
jet axis is around the jet axis (position 0), as can be expected.  We
find that the largest radial velocities are found in the OP jet with velocities 
above 100 km/s, and gradually lower velocities for more negative $\mu$. 
In the PP jet the maximum perpendicular velocity is found to be about 60
km/s at 15 AU. The maximum velocity found
does not differ much between the lines.

\subsection{Jet rotation far from the source}
\label{rotationsection}

In \citet{staff10} we found that the Keplerian rotational profile in the
disc is inherited by the jet, and is clearly visible in the jet at least 30
AU from the disc (the limitations were due to the size of the simulation
box).  This applied both to the OP and BP jet simulations studied there. 
For the OP jet, only the inner jet exhibited this feature.  We again find
the same features in this work.  The $\mu=-0.12$ simulation which also has a
two-component jet structure also shows the Keplerian-like rotation profile
for the inner jet (we have not attempted to fit it with a Keplerian rotation
profile).  The PP jet, on the other hand, is different in that the kink mode
is visible throughout most of the jet
(this can be seen as a
corkscrew pattern in the PP jet in Figs.~\ref{imageClIV} and~\ref{imageSII},
and also to some extent in Fig.~\ref{imageMgII}).
This kink mode is visible in all the
jets closer to the head of the jet (far out). In the PP jet the kink mode
does not appear to get out of control, while in the other jets the spiral
appears to grow farther out in the jet. In the analysis of
\citet{pelletier92}, it was shown that the PP magnetic configuration
characterizes a minimum energy state for jets which may explain why this
model appears to be more stable than the other three.

\begin{figure*}
\includegraphics[width=\textwidth]{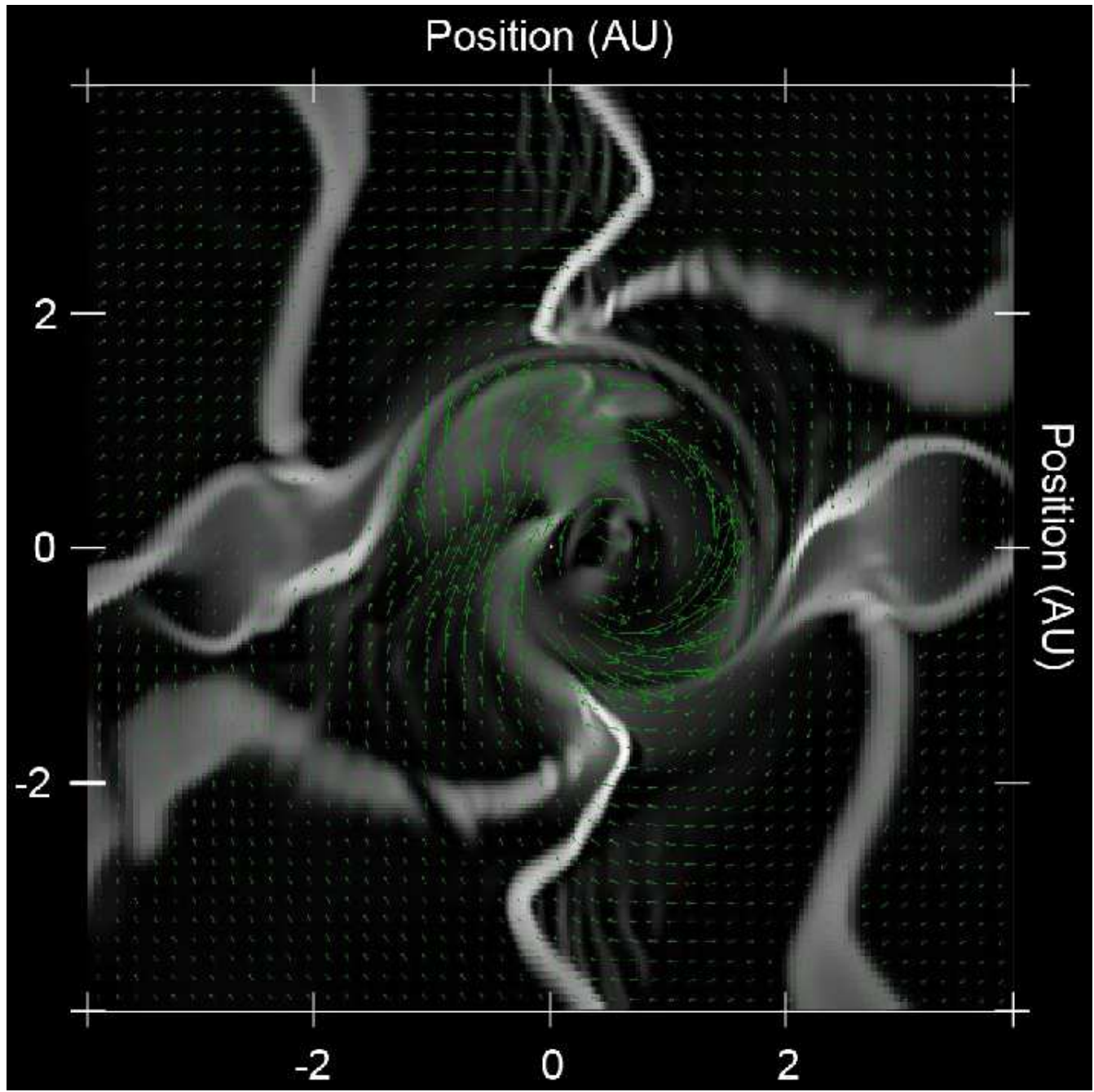}
\caption{A slice through the PP jet taken perpendicular to the jet 24 AU
from the disc. Only the innermost $\pm4$ AU is plotted for clarity. Shown is 
filled density contours with the green vectors 
illustrating the velocity components in that plane. The jet can undergo
counter-rotation, and in this particular slice it even appears
as if the jet is rotating ``both ways'', which is an artefact of the MHD
kink (m=1) mode at play. }
\label{PPvelvecden}
\end{figure*}

Figure~\ref{PPvelvecden} shows a cut perpendicular to the jet axis 24 AU
from the disc of the density and velocity in the PP jet. At this particular 
cut it appears as if part of the jet is rotating one direction, and another part
is rotating in the opposite direction.
This rotation is also off centered. On this basis, we argue that this is
why \citet{coffey12} had difficulty finding any indication of
rotation in the RW Aur jet\footnote{Animations ``scanning'' through the jet, from the
disc outwards, for the BP and PP simulations, can be viewed here:\\
\texttt{http://quarknova.ucalgary.ca/MHD.html}
\\The evolution of the density and the magnetic field lines for all
the simulations can also be viewed.}. 
In the previous section we showed that the RW Aur jet is rather narrow, and
therefore more similar to the more negative $\mu$ jets (BP and PP). This is
an independent indication that the RW Aur jet may have a steeper power law
dependence of the magnetic field (more negative $\mu$)
on the disc, that can cause counter rotation and
possibly explain the conflicting observations of rotation in this jet.

\begin{figure*}
\includegraphics[width=\textwidth]{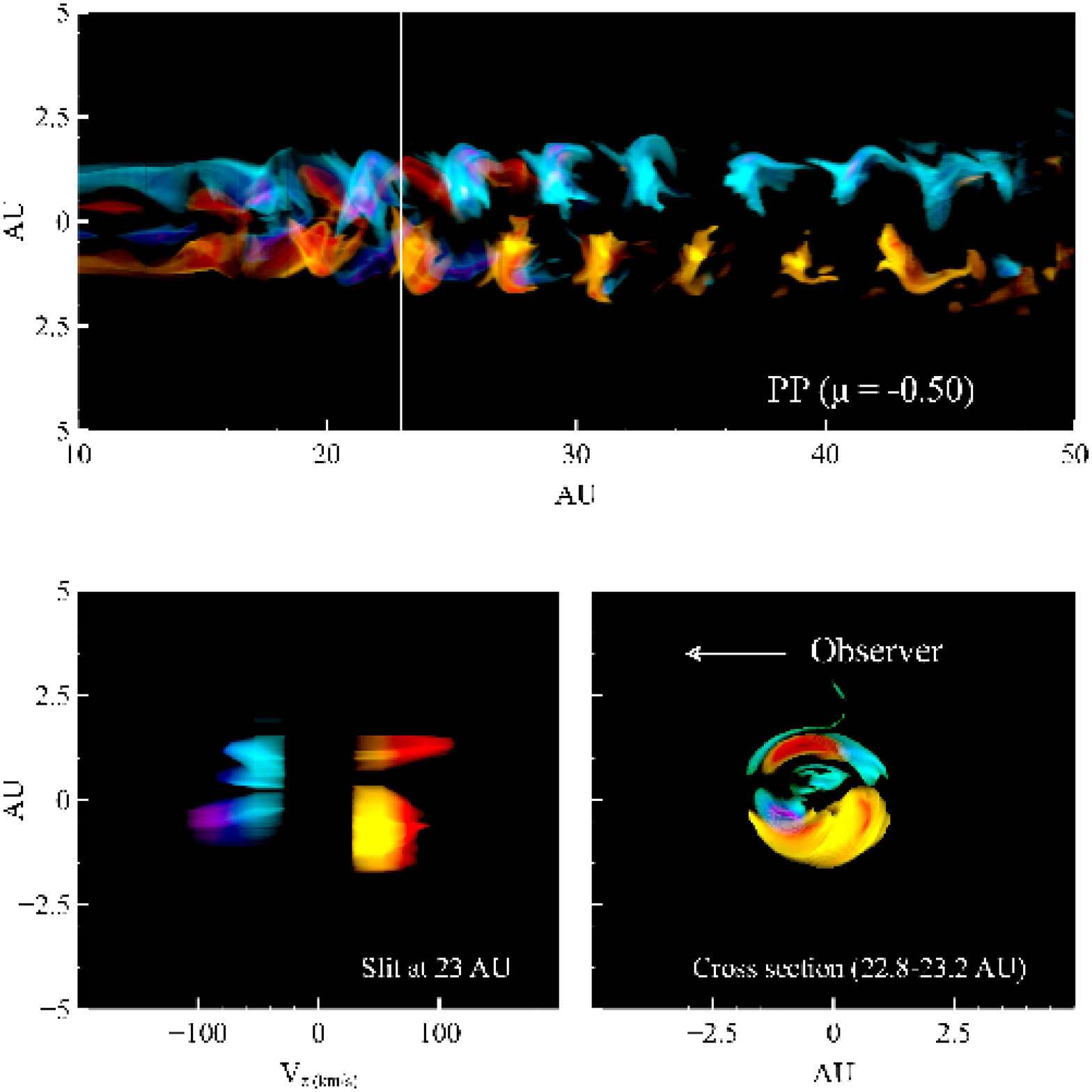}
\caption{Demonstration of the two rotational components in the PP jet. The
top panel shows the rotational velocity of the inner part of the jet.
Blueish colors designate gas moving away from us, yellow and redish colors
gas coming towards us. The disc rotates clock-wise, that is away from us
above the axis and towards us below the axis in the top panel. From the very 
early phases of the jet we see that
there is a thin component rotating opposite of the main jet, ie. towards us
(red color) above the axis, and away from us (blue color) below the axis. From
$\sim16~{\rm AU}$ the spiral occurs, which is this inner component. We make
a cut perpendicular to the jet at 23 AU (illustrated with a white line),
where we make the position velocity diagram seen in the lower left panel.
The lower right panel shows the rotational velocity in this perpendicular
slice. }
\label{pprotation}
\end{figure*}

In the PP configuration with a high magnetic flux close to ${\rm r_i}$,
matter is very efficiently drained from the region around the axis close to
the protostar (within $\sim1~{\rm r_i}$).  This is the region where no
matter is injected from the disc.  The resulting extremely low density
region slows down the simulations drastically and we remedy to this by
enforcing a floor density which effectively consist of replenishing the
region with matter without angular momentum.  
In all our simulations the gas around the axis near the star 
is flowing towards the star (back-flow), but only in the PP simulation does 
it change into an outflow.  Only a very small region is affected by this,
and since we do not draw vectors for all grid cells in Fig.~\ref{momentum},
this can not be seen in that figure.

This low density outflow is accellerated to large velocity, soon reaching
$M_{\rm A}>1$.  The kink mode instability which can occur when $M_{\rm A}>1$
\citep{diaz11}, is clearly prominent in the PP case and starts very close to
the disc.  Furthermore, the strong (i.e.  low $\beta$) axial magnetic field
along the back-bone (i.e.  in the region within $r_i$) can sustain and
confine
 a forward non-rotating outflow (in the + z-direction). This non-rotating
flow is channeled along the back-bone and results from the magnetic
"squeezing" of the low-density coronal gas within $r_i$ due to the strong
$B_{\phi}$.  The $m=-1$ mode grows and is also dynamically important in the
PP case in the vicinity of $r_i$ and close to the disk.  This complex
dynamics leads, in the case of PP, to the rotating and counter-rotating
components (and a forward component) as shown in Fig.~\ref{pprotation}.

The extra mass added on the axis is crucial in order to get
counter-rotation, since without it there would not be any matter that could
counter-rotate.  In reality, this mass could be provided by a wind from the
protostar \citep[i.e.][]{matt05}.  The back-flow remains throughout the
simulations in all but the PP configuration, and this is why it is difficult
to find a counter-rotating component in those configurations.  A stellar
wind would have to be sufficiently strong to overcome the ram pressure of
this back-flow in order to get anywhere.  That is, only a strong stellar
wind could do it.  But a strong stellar wind alone will not give
counter-rotation.  It is also necessary to have twisted field lines along
the axis, if not a spiral, being twisted in the direction opposite of the
disc rotation that the wind can flow along to cause counter-rotation. 
However, in order for the gas to follow the field lines rather than just
dragging the field with the flow, the magnetic pressure must be greater than
the ram pressure of the wind.  Hence the wind can not be too strong.

The gas in the jet that is not part of the spiral continues to rotate
clockwise, in the same direction as the disc. In some sense, the PP jet also
has 2 components, though of a very different nature than the 2 component OP
and $\mu=-0.12$ jets. In the PP jet, one component counter-rotates, the
other co-rotates with the disc as illustrated in Fig.~\ref{pprotation}.
It is challenging at this stage to assess exactly how much of the jet is
counter-rotating.  Nevertheless, rough estimate can be derived by looking at
Fig.~\ref{pprotation}.  We find that between 10 and 50 AU, we observe about
$10\%$ counter-rotation.  This is not to say that $10\%$ of the mass in the
jet is counter-rotating.  In fact very little mass is counter-rotating, but
it is very hot, making it emit strongly.  The likelihood of detecting this
depends on how well the jet is observed.  The counter-rotating part is
reasonably bright, so if more of the fainter gas surrounding the spiral is
detected, less counter-rotation should be seen. The counter-rotation becomes
less pronounced farther out in the jet, and beyond about 50 AU we do not
find much counter-rotation. At this point the front of the jet is at 90 AU,
and we expect that as the jet develops further, the counter-rotating part
will also extend further.
We note that this is quite different from the
counter-rotation found in an analytic study by \citet{sauty12}.

 Whether the observed velocity gradients across the jet are hints of rotation \citep{pech12}
or  are a result of sub-jets structure \citep{soker12} remains to be confirmed. 
In general, measuring  rotation in jets is a challenging exercise 
\citep[e.g.][]{coffey12}. 
 From the analysis of our simulations we find that many effects can be
 misinterpreted as rotation.  Only by looking at the jet simultaneously in
 different lines and in PV diagrams can we distinguish clean signs of
 rotation from velocity gradients induced by MHD modes and those induced by
 the two-component jets.

\subsection{Fluxes and jet energetics}

\begin{figure*}
\centering
\includegraphics[width=0.7\textwidth]{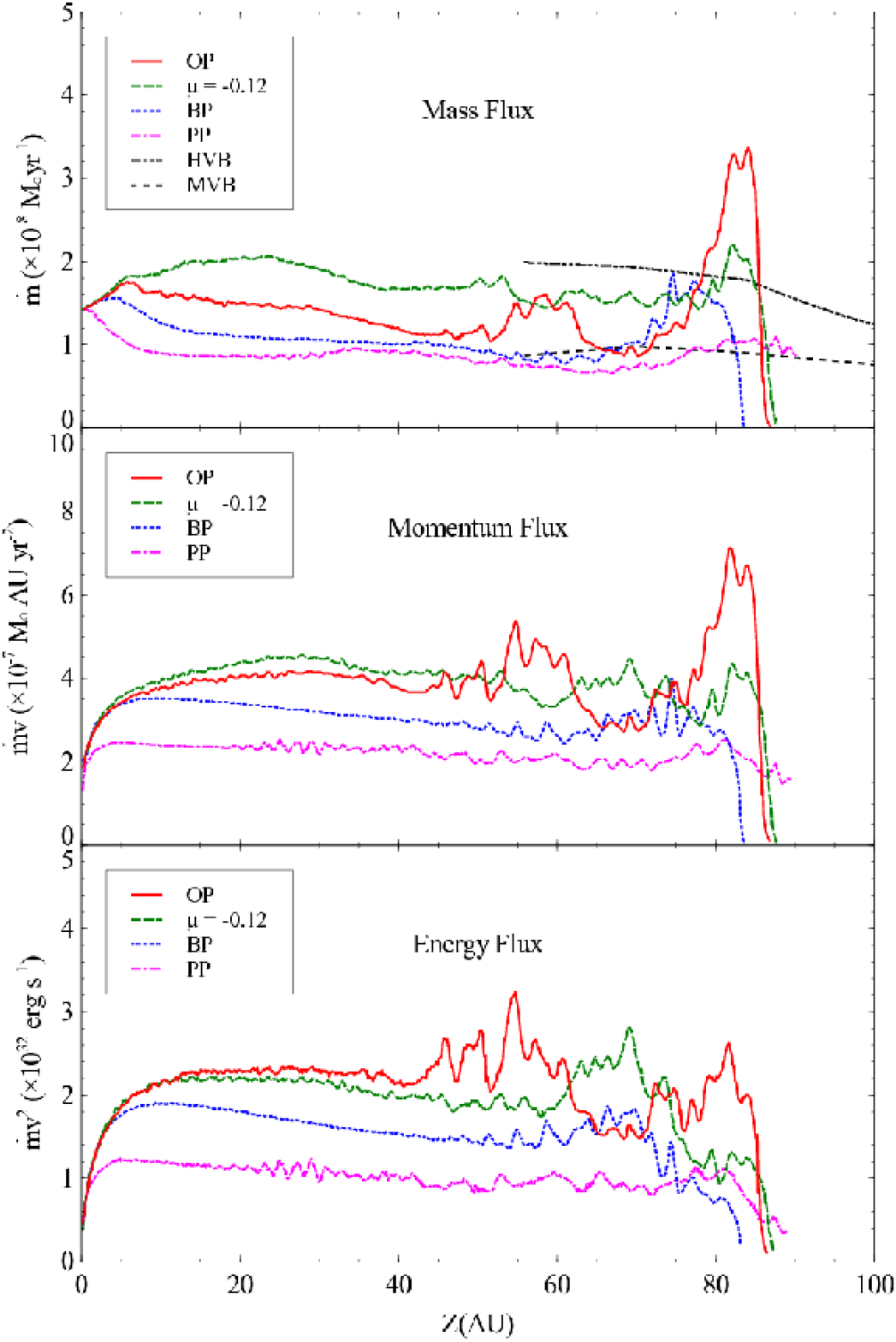}
\caption{Mass flux (top panel), momentum flux (middle panel), and jet power
(bottom panel) along the jet. Red is OP, green is $\mu=-0.12$, blue is 
BP, and purple is PP. In the mass flux figure, we also overplot the observed
mass flux in DG tau from \citet{agra-amboage11} for the high (HVB) and 
medium velocity (MVB)
blueshifted intervals. The $\mu=-0.12$ jet has the highest mass flux, but
because the OP jet is faster, the two have similar momentum and the OP jet
has the higher jet power. The BP and PP jets are generally at lower fluxes
throughout the length of the jet. }
\label{flux}
\end{figure*}

Figure~\ref{flux} shows the mass flux, the momentum flux, and the power in
the jets.  The mass flux for the OP and BP jet can be compared to the mass
flux reported in \citet{staff10}.  We now find that both the OP and the BP
flux is slightly higher than in \citet{staff10}, with the difference due to
wider jets in the larger simulations. The jet
appears to have expanded sideways when propagating from 60 AU to 90 AU. 
In contrast to \citet{staff10}, we have simply used a constant jet width along
the entire jet, this width is given in Table~\ref{massfluxtable}.  As in
\citet{staff10}, we find that the mass flux in the OP jet is larger than in
the BP jet.  The $\mu=-0.12$ jet has an even higher mass flux than the OP
jet, only in the very front of the jet has the OP jet higher mass flux
indicating that it has piled up more matter in the front. This could in part
be because the OP jet has a less pointy front because of the lower Mach
number. The mass flux
in the PP jet is lower than the mass flux in the BP jet throughout.
The observed mass flux in the DG tau jet \citep[taken from][]{agra-amboage11} 
is also shown in Fig.~\ref{flux}. The mass
fluxes that we find are comparable to those in observed jets
\citep[e.g.][for the DG tau jet]{agra-amboage11}.
Because the OP jet is faster, the momentum flux in the OP and the
$\mu=-0.12$ jets are similar, and the OP jet power is larger.

\begin{table*}
\centering
\caption{General Properties of simulated jets.}
\label{massfluxtable}
\begin{tabular}{ccccc}\\\hline
Case & OP ($\mu=-0.01$) & $\mu=-0.12$ & BP ($\mu =-0.25$) & PP ($\mu =-0.5$) \\\hline
Two-components & Yes & Yes & Weak & No \\\hline
Inner-jet opening angle$^\ddagger$ &  7.0 & 5.6 & recollimation$^*$ &
recollimation$^*$ \\\hline
full width$^\dagger$ & $25$ AU & $30$ AU & $20$ AU$^{**}$ & $20$ AU \\
\hline
\end{tabular}\\
{\raggedright 
$^\ddagger$ The jet opening angle for the OP and the $\mu=-0.12$ jets, which
remains fairly constant.\\
$^*$ In the BP and PP simulations the jet recollimates and therefore we do
not define an opening angle.
\\
$^{**}$ Ignoring the weak outer jet.
\\
$^\dagger$ The maximum jet width for the end state for each 
magnetic field configuration simulated.
}
\end{table*}

\subsection{Justification of the non-uniform gird: Numerical
verification}

As outlined in section~\ref{numericalsection}, we use a uniform grid of 100
cells in the $x_2$ and $x_3$ directions only between $-25 r_i$ and $25 r_i$.
Beyond this, we use a ratioed grid, in order to stretch the grid size and
thereby contain the jet on the grid. Likewise, only 200 cells in the $x_1$
direction between $0$ and $100 r_i$ is in a uniform grid, with a ratioed
grid beyond that. We have verified that the ratioed grid does not appear to
affect the results by making a much smaller grid with 980x660x660 cells in a
uniform grid extending from $0$ to $490 r_i$ in the $x_1$ direction, and
from $-165 r_i$ to $165 r_i$ in the $x_2$ and $x_3$ directions. In the early
phases of the jet evolution, while the jet remains on this smaller grid, we
find the same results as on the bigger grid. 

In particular, the 2-component jet structure in the OP simulation is also
apparent at this early phase of the jet evolution. Because of this, we
conclude that the existence of a 2 component jet structure is not related 
to the non-uniform grid. At this early stage, the outer jet is also far from
any of the boundaries, in particular in the larger simulation box with the
ratioed grid, and we conclude that the existence of the outer jet is
unrelated to the boundaries of the simulation box.

\section{Summary}

We have presented the results of four 3D MHD simulations of protostellar
jets, with different initial magnetic field structure.  The jets are
followed (simulated) from the source, and the simulation box
stretches out to 90 AU along the jet, and 27 AU on either side of the jet
axis making the simulation box 90 AU x 54 AU x 54 AU in size.  

In order to better compare our results with observations, we have created
synthetic emission line maps and PV diagrams. We focused on 
[SII] $6730~ {\rm \AA}$ and 
MgII $2796~ {\rm \AA}$ as these have commonly been used to observe
protostellar jets. In addition, we also created synthetic emission line maps
in [ClIV] $74500~{\rm \AA}$ which has a much lower critical density than
the [SII] and MgII lines, in order to illustrate the effect of different
critical density regimes.
We have used these synthetic lines to perform our analysis.
The initially more open magnetic field configuration (the PP model), results
in the most collimated jet (even when ignoring the possible outer jet in
other simulations), with a strong recollimation about halfway
through the simulation box. This feature is visible also in the magnetic
field lines. The inner jet in the OP and $\mu=-0.12$ simulations appear to
have a constant opening angle of about 7 degrees (OP), and 5.6 degrees
($\mu=-0.12$).

We confirm the results from \citet{staff10} regarding jet rotation, that the
inner jet in the OP and $\mu=-0.12$ simulations has a Keplerian rotation
profile. However, this only extends out a few AU from the rotation axis,
then the Keplerian rotation profile is broken by the faster rotating outer
jet. The BP jet shows a clear Keplerian profile throughout (although there is 
a slight deviation in
the rotation profile associated with the weak outer jet).  We find that the
PP jet is different, in that the kink mode is visible throughout most of the
jet. This leads to regions in the jet that are counter-rotating. 
Hence, memory of the underlying disc Kepler profile is not always preserved. 
{\it We emphasize that not observing a Kepler profile does not mean it is
not a centrifugal wind.  } We also find this corkscrew jet in the outer part
of the other simulations, and while the corkscrew feature seems to grow
farther from the disc in these simulations, it appears to remain with a
constant amplitude in the PP simulation.

Perhaps one of the most important results of our simulations is that it is
possible to produce counter-rotating regions in jets as a consequence of
kink instabilities. 
That may explain the observations of counter-rotation in jets such as
the RW Aur jet.

Laboratory studies of magnetized plasma jets provide replica of
astrophysical jets \citep[e.g][]{hsu02,ampleford08,frank09}.
In these laboratory jets, the kink mode and spiral-like
structure seem to develop in the laboratory jets with similarities with our
findings.  In \citet{hsu02}, the magnetic field on the ``disc'' appears even
more open than the PP field which could explain their extreme geometry. 
However since the ``disc'' (the annulus) in these experiments does not   
effectively rotate the origin of the spiral is harder to understand.  In
laboratory studies
 which include rotation \citep{ampleford08} the magnetic field did not
play a significant dynamical role unlike what we found in our simulations.
Finally, \citet{frank09} found that the MHD (kink mode and/or sausage)
instabilities in jets break
 the laboratory plasma jets to into ``chunks" reminiscent of the knotty
features we see in the OP jet.  The many similarities between laboratory
jets and astrophysical jets despite the immense difference in scale speaks
to the fundamental nature of the MHD jets.

We still lack a robust explanation for the physics behind the transition
from a one- to two-component jet; i.e.  how to translate $\mu$ into a true
physical mechanism that leads to the formation of the outer jet.  For now we
can only speculate based on a few facts we observed when comparing the
launching, formation and collimation of the jet in the simulations.  In the
OP and $\mu=-0.12$ simulations, because of the more narrowly opened magnetic
field lines along the disc, the inner disc region participating in a wind is
larger than in the PP and BP cases.  Furthermore, the strong back-flow seems
to split the jet into two components.  The PP  configuration do not
seem to develop a clear outer jet.  The material flying out sideways in the
PP simulations ends up digging out a bigger cavity around the narrow axial
jet.  The BP jet appears as an intermediate case, with a very weak outer
jet.

There are hints of  two-component jets   at different scales including AGNs
\citep{asada10}
 and maybe in GRBs \citep{filgas11}.  This may be an indication, as our simulations suggest,  that the
 $\mu$ effect is real and may operates at different scales. This is not an unreasonable assumption given the
  universal nature   of astrophysical jets.  
We need to perform more simulations
 by varying $\mu$ to assess if there is  a critical $\mu$ which separates
the one-component jets from the two-component jets.  If such a critical
value exists it is most likely
 around $\mu= -0.25$.   What is evident is the
  fact that the strength of the outer jet (i.e. for simulations with $\mu
\ge 0.25$)
   increases with $\mu$.  For example, we find the outer jet to rotate
slower in the $\mu=-0.12$ case than in the OP ($\mu=-0.01$).

These MHD simulations of large scale disc winds lead to features similar to
those of observed low mass protostellar jets.  The jets reach a maximum
width of 20-30 AU, comparable to the width of observed YSO jets.  We find
the mass flux in the jet to be of the order $1-2\times10^{-8} M_\odot~{\rm
yr}^{-1}$ (comparable to the observed mass flux in the DG tau jet), with the
more open disc magnetic field configurations (BP and PP) having a somewhat
lower mass flux.  However, the $\mu=-0.12$ jet has a larger mass flux than
the least open disc field configuration (OP), as it is wider.  Only in the
head of the jet does the OP jet carry more mass than the $\mu=-0.12$ jet.  
We have also found that the inner jet preserves the underlying Keplerian
rotation profile to large distance. However, for the most open disc magnetic
field configuration, the kink mode creates a
corkscrew-like jet without a clear Keplerian rotation profile. In this case,
we even find regions in the jet rotating opposite to the disc.
No outer jet develops in this case. We conclude that magnetized disc
winds from underlying Keplerian discs can develop rotation profiles far down
the jet that are not Keplerian, or even counter-rotating owing to the
operation of kink modes in the jet.

\section*{Acknowledgements}
We thank the anonymous referee for helpful remarks. We thank J. Ge for help with  Fig.~\ref{visitfigures}.
This work was made possible by the facilities of the Shared Hierarchical 
Academic Research Computing Network (SHARCNET:www.sharcnet.ca) and
Compute/Calcul Canada.
J.E.S acknowledges support from the Australian Research Council Discovery 
Project (DP12013337) program.
This work has been supported, in part, by grant AST-0708551
from the U.S. National Science Foundation,
and, in part, by grant NNX10AC72G from NASA's ATP program.
RO and REP  are supported by the Natural Sciences and Engineering
Research Council of Canada.


\end{document}